
%

\def\ie{{\it i.e.}}

\def\etal{et~al.}

%
\def\arev #1 #2 {ARA\&A~{#1}, {#2}}
\def\aj #1 #2 {AJ~{#1}, {#2}}
\def\aa #1 #2 {A\&A~{#1}, {#2}}
\def\aas #1 #2 {A\&AS~{#1}, {#2}}
\def\aal #1 #2 {A\&A~{#1}, L{#2}}
\def\aar #1 #2 {A\&AR~{#1}, {#2}}
\def\apj #1 #2 {ApJ~{#1}, {#2}}
\def\apjs #1 #2 {ApJS~{#1}, {#2}}
\def\apjl #1 #2 {ApJ~{#1}, L{#2}}
\def\apss #1 #2 {Ap\&SS {#1}, {#2}}
\def\baas #1 #2 {BAAS~{#1}, {#2}}
\def\mn #1 #2 {MNRAS~{#1}, {#2}}
\def\mnras #1 #2 {MNRAS~{#1}, {#2}}
\def\nat #1 #2 {Nat~{#1}, {#2}}
\def\nature #1 #2 {Nat~{#1}, {#2}}
\def\pasj #1 #2 {PASJ~{#1}, {#2}}
\def\pasp #1 #2 {PASP~{#1}, {#2}}
\def\sov #1 #2 {Sov.Astron.~{#1}, {#2}}
\def\sovl #1 #2 {Sov.Astron.~{#1}, {#2}}
\def\asr #1 #2 {Adv. in Space Research~{#1}, {#2}}
\def\ssr #1 #2 {Space Sci. Rev.~{#1}, {#2}}
\def\physrevd #1 #2 {Phys. Rev. D~{#1}, {#2}}
\def\iauc #1 #2 {IAU Circ.~{#1}}
%
\def\smallcap{\relax}
\def\sc{\relax}
\font\sevbf=cmbx7
\font\eightrm=cmr8
\font\eightbf=cmbx8
\font\ninerm=cmr9
\font\ninei=cmmi9
\font\nineit=cmti9
\font\ninesl=cmsl9
\font\ninebf=cmbx9
\font\ninesy=cmsy9
\def\rmnine{\fam0\ninerm}
\def\itnine{\fam\itfam\nineit}
\def\slnine{\fam\slfam\ninesl}
\def\bfnine{\fam\bffam\ninebf}
\def\ninepoint{\let\rm=\rmnine
\textfont0=\ninerm\scriptfont0=\sevenrm
\textfont1=\ninei\scriptfont1=\seveni
\textfont2=\ninesy
\textfont\itfam=\nineit \let\it=\itnine
\textfont\slfam=\ninesl \let\sl=\slnine
\textfont\bffam=\ninebf \scriptfont\bffam=\sevbf
\let\bf=\bfnine
\let\smallcap=\sevenrm
\let\sc=\sevenrm
\normalbaselineskip=10pt\normalbaselines\rm}
\font\tenib=cmmib10
\font\tenslbf=cmbxsl10

\def\rmten{\fam0\tenrm}
\def\itten{\fam\itfam\tenit}
\def\slten{\fam\slfam\tensl}
\def\bften{\fam\bffam\tenbf}
\def\tenpoint{\let\rm=\rmten
\textfont0=\tenrm\scriptfont0=\sevenrm
\textfont1=\teni\scriptfont1=\seveni
\textfont2=\tensy
\textfont\itfam=\tenit \let\it=\itten
\textfont\slfam=\tensl \let\sl=\slten
\textfont\bffam=\tenbf \scriptfont\bffam=\ninebf
\let\bf=\bften
\let\smallcap=\eightrm
\let\sc=\eightrm
\normalbaselineskip=12pt\normalbaselines\rm}
\def\rmbften{\fam0\tenbf}
\def\itbften{\fam\itfam\tenib}
\def\slbften{\fam\slfam\tenslbf}
\def\bftenpoint{\let\rm=\rmbften
\textfont0=\tenbf\scriptfont0=\sevbf
\textfont1=\tenib\scriptfont1=\seveni
\textfont2=\tensy
\textfont\itfam=\tenib \let\it=\itbften
\textfont\slfam=\tenslbf \let\sl=\slbften
\textfont\bffam=\tenbf \scriptfont\bffam=\ninebf
\let\bf=\bften
\let\smallcap=\eightbf
\let\sc=\eightbf
\normalbaselineskip=12pt\normalbaselines\rm}
\font\twelbf=cmb10 scaled \magstep1
\font\twelmib=cmmib10 scaled \magstep1
\font\twelslbf=cmbxsl10 scaled \magstep1
\font\fortbf=cmb10   scaled \magstep2
\font\fortsy=cmsy10   scaled \magstep2
\def\rmbffort{\fam0\fortbf}
\def\itbffort{\fam\itfam\twelmib}
\def\slbffort{\fam\slfam\twelslbf}
\def\bffort{\fam\bffam\fortbf}
\def\bffortpoint{\let\rm=\rmbffort
\textfont0=\fortbf  \scriptfont0=\twelbf
\textfont1=\twelmib \scriptfont1=\twelmib
\textfont2=\fortsy
\textfont\itfam=\twelmib  \let\it=\itbffort
\textfont\slfam=\twelslbf \let\sl=\slbffort
\textfont\bffam=\fortbf \scriptfont\bffam=\twelbf
\let\bf=\bffort
\let\smallcap=\twelbf
\let\sc=\twelbf
\normalbaselineskip=16pt\normalbaselines\rm}
%

\vsize=23.5truecm
\hoffset=-1true cm
\voffset=-1true cm
\newdimen\fullhsize
\fullhsize=40cc
\hsize=19.5cc
\def\fullline{\hbox to\fullhsize}
\def\makefootline{\baselineskip=10dd \fullline{\the\footline}}
\def\makeheadline{\vbox to 0pt{\vskip-22.5pt
            \fullline{\vbox to 8.5pt{}\the\headline}\vss}\nointerlineskip}
\let\lr=L \newbox\leftcolumn
\output={\global\topskip=10pt
         \if L\lr
            \global\setbox\leftcolumn=\columnbox \global\let\lr=R
            \message{[left\the\pageno]}%
            \ifnum\pageno=1
               \global\topskip=\fullhead\fi
         \else
            \doubleformat \global\let\lr=L
         \fi
         \ifnum\outputpenalty>-2000 \else\dosupereject\fi}
\def\doubleformat{\shipout\vbox{\makeheadline
    \fullline{\box\leftcolumn\hfil\columnbox}
           \makefootline}
           \advancepageno}
\def\columnbox{\leftline{\pagebody}}
\outer\def\bye{\sterne=1\ifx\speciali\undefined\else
\loop\smallskip\noindent special character No\number\sterne:
\csname special\romannumeral\sterne\endcsname
\advance\sterne by 1\global\sterne=\sterne
\ifnum\sterne<11\repeat\fi
\if R\lr\null\fi\vfill\supereject\end}
\hfuzz=2pt
\vfuzz=2pt
\tolerance=1000
\fontdimen3\tenrm=1.5\fontdimen3\tenrm
\fontdimen7\tenrm=1.5\fontdimen7\tenrm
\abovedisplayskip=3 mm plus6pt minus 4pt
\belowdisplayskip=3 mm plus6pt minus 4pt
\abovedisplayshortskip=0mm plus6pt
\belowdisplayshortskip=2 mm plus4pt minus 4pt
\predisplaypenalty=0
\clubpenalty=20000
\widowpenalty=20000
\parindent=1.5em
\frenchspacing
\def\newline{\hfill\break}%

\newtoks\vol
\newtoks\startpage \newtoks\lastpage
\newtoks\month \newtoks\year
\vol={1} %
\month={JANVIER}\year={2001}%
\startpage={1}\lastpage={99}%
\nopagenumbers
\def\paglay{\headline={
{\ninepoint\hsize\fullhsize\ifnum\pageno=1
\vtop{\baselineskip=10dd\hrule width52.5mm height0.4pt \kern5pt
\hbox{\noindent ASTRONOMY\ \ \&\ \ ASTROPHYSICS}\kern5pt
\hbox{SUPPLEMENT\ \ SERIES}
\kern5pt\hrule width52.5mm height0.4pt \kern5pt
\hbox{{\sl Astron. Astrophys. Suppl. Ser.}
{\bf \the\vol ,}\ \ \the\startpage -\the\lastpage\ \ (\the\year)}}
\hfill\vtop{\kern5pt\hbox{\the\month\ {\bf \the\year ,} PAGE \the\startpage}}
\else\ifodd\pageno --- \hfil--- --- ---\hfil\folio
\else \folio\hfil --- --- --- \hfil---\fi\fi}}}
\ifx \undefined\instruct
\headline={\tenrm\ifodd\pageno\hfil\folio
\else\folio\hfil\fi}\fi
\newcount\eqnum\eqnum=0
\def\autnum{\global\advance\eqnum by 1{\rm(\the\eqnum)}}

\def\utw{\smash{\rlap{\lower5pt\hbox{$\sim$}}}}
\def\udtw{\smash{\rlap{\lower6pt\hbox{$\approx$}}}}

\def\diameter{{\ifmmode\mathchoice
{\ooalign{\hfil\hbox{$\displaystyle/$}\hfil\crcr
{\hbox{$\displaystyle\mathchar"20D$}}}}
{\ooalign{\hfil\hbox{$\textstyle/$}\hfil\crcr
{\hbox{$\textstyle\mathchar"20D$}}}}
{\ooalign{\hfil\hbox{$\scriptstyle/$}\hfil\crcr
{\hbox{$\scriptstyle\mathchar"20D$}}}}
{\ooalign{\hfil\hbox{$\scriptscriptstyle/$}\hfil\crcr
{\hbox{$\scriptscriptstyle\mathchar"20D$}}}}
\else{\ooalign{\hfil/\hfil\crcr\mathhexbox20D}}%
\fi}}
\normalbaselines\rm
\def\petit{\def\rm{\fam0\eightrm}
\textfont0=\eightrm \scriptfont0=\sixrm \scriptscriptfont0=\fiverm
 \textfont1=\eighti \scriptfont1=\sixi \scriptscriptfont1=\fivei
 \textfont2=\eightsy \scriptfont2=\sixsy \scriptscriptfont2=\fivesy
 \def\it{\fam\itfam\eightit}%
 \textfont\itfam=\eightit
 \def\sl{\fam\slfam\eightsl}%
 \textfont\slfam=\eightsl
 \def\bf{\fam\bffam\eightbf}%
 \textfont\bffam=\eightbf \scriptfont\bffam=\sixbf
 \scriptscriptfont\bffam=\fivebf
 \def\tt{\fam\ttfam\eighttt}%
 \textfont\ttfam=\eighttt
 \let\tams=\kleinhalbcurs
 \let\tenbf=\eightbf
 \let\sevenbf=\sixbf
 \normalbaselineskip=9dd
 \if Y\REFERE \normalbaselineskip=2\normalbaselineskip
 \normallineskip=2\normallineskip\fi
 \setbox\strutbox=\hbox{\vrule height7pt depth2pt width0pt}%
 \normalbaselines\rm}%
\def\begpet{\vskip6pt\bgroup\ninepoint}
\def\endpet{\vskip6pt\egroup}
\newbox\FigT
\def\rahmen#1{
\setbox\FigT=\vbox{\hbox to4true cm{
\hfil\vrule width 0.01pt height 2cm\hfil}
\hrule height 0.01pt}
\vbox to#1true cm{\vfil\line{\hfill\box\FigT\hfill}}}
\def\begfig#1cm#2\endfig{\par
   \ifvoid\topins\midinsert\bigskip\vbox{\rahmen{#1}#2}\endinsert
   \else\topinsert\vbox{\rahmen{#1}#2}\endinsert
\fi}
\def\begfigwid#1cm#2\endfig{\par
\if N\lr\else
\if R\lr
\shipout\vbox{\makeheadline
\line{\box\leftcolumn}\makefootline}\advancepageno
\fi\let\lr=N
\topskip=10pt
\output={\plainoutput}%
\fi
\topinsert\line{\vbox{\hsize=\fullhsize\rahmen{#1}#2}\hss}\endinsert}
\def\figure#1#2{\bigskip\noindent{\ninepoint F{\sc IGURE} #1.\
\ignorespaces #2\smallskip}}
\def\begtab#1cm#2\endtab{\par
   \ifvoid\topins\midinsert\medskip\vbox{#2\rahmen{#1}}\endinsert
   \else\topinsert\vbox{#2\rahmen{#1}}\endinsert
\fi}
\def\begtabemptywid#1cm#2\endtab{\par
\if N\lr\else
\if R\lr
\shipout\vbox{\makeheadline
\line{\box\leftcolumn}\makefootline}\advancepageno
\fi\let\lr=N
\topskip=10pt
\output={\plainoutput}%
\fi
\topinsert\line{\vbox{\hsize=\fullhsize#2\rahmen{#1}}\hss}\endinsert}
\def\begtabfullwid#1\endtab{\par
\if N\lr\else
\if R\lr
\shipout\vbox{\makeheadline
\line{\box\leftcolumn}\makefootline}\advancepageno
\fi\let\lr=N
\output={\plainoutput}%
\fi
\topinsert\line{\vbox{\hsize=\fullhsize\noindent#1}\hss}\endinsert}

\def\begfullpage{\vfill\supereject
            \if R\lr\null\vfill\supereject\fi
            \begingroup\output={\plainoutput}
            \hsize=\fullhsize}
\def\endfullpage{\vfill\supereject\endgroup\let\lr=L}
\def\begref{\begingroup\let\INS=N\let\refer=\ref}
\def\ref{\goodbreak\if N\INS\let\INS=Y\vbox{\bigskip\noindent\tenbf
References\bigskip}\fi\hangindent\parindent
\hangafter=1\noindent\ignorespaces}
\def\endref{\endgroup}
\def\ack#1{\vskip11pt\begingroup\noindent{\bf Acknowledgements\/}.
\ignorespaces\vskip11pt\noindent#1\vskip6pt\endgroup}

%
 \def \aTa  { \goodbreak
     \bgroup
     \par
     \rightskip=0pt plus2cm\spaceskip=.3333em \xspaceskip=.5em
     \pretolerance=10000
     \noindent
     \bffortpoint}
 %
 \def \eTa{\vskip10pt\egroup
     \noindent
     \ignorespaces}
%
 \def \aTb{\goodbreak
     \bgroup
     \par
     \rightskip=0pt plus2cm\spaceskip=.3333em \xspaceskip=.5em
     \pretolerance=10000
     \noindent
     \bftenpoint}
 %
 \def \eTb{\vskip10pt
     \egroup
     \noindent
     \ignorespaces}
%
\catcode`\@=11
\expandafter \newcount \csname c@Tl\endcsname
    \csname c@Tl\endcsname=0
\expandafter \newcount \csname c@Tm\endcsname
    \csname c@Tm\endcsname=0
\expandafter \newcount \csname c@Tn\endcsname
    \csname c@Tn\endcsname=0
\expandafter \newcount \csname c@To\endcsname
    \csname c@To\endcsname=0
\expandafter \newcount \csname c@Tp\endcsname
    \csname c@Tp\endcsname=0
\def \resetcount#1    {\global
    \csname c@#1\endcsname=0}
\def\@nameuse#1{\csname #1\endcsname}
\def\arabic#1{\@arabic{\@nameuse{c@#1}}}
\def\@arabic#1{\ifnum #1>0 \number #1\fi}
 %
\expandafter \newcount \csname c@fn\endcsname
    \csname c@fn\endcsname=0
\def \stepc#1    {\global
    \expandafter
    \advance
    \csname c@#1\endcsname by 1}
\catcode`\@=12
%
%
   \catcode`\@= 11
%
\def\footnoterule{\kern-3pt\hrule width 2true cm\kern2.6pt}
\newinsert\footins
\def\footnotea#1{\let\@sf\empty 
  \ifhmode\edef\@sf{\spacefactor\the\spacefactor}\/\fi
  {#1}\@sf\vfootnote{#1}}
\def\vfootnote#1{\insert\footins\bgroup
  \ninepoint
  \interlinepenalty\interfootnotelinepenalty
  \splittopskip\ht\strutbox 
  \splitmaxdepth\dp\strutbox \floatingpenalty\@MM
  \leftskip\z@skip \rightskip\z@skip \spaceskip\z@skip \xspaceskip\z@skip
  \textindent{#1}\footstrut\futurelet\next\fo@t}
\def\fo@t{\ifcat\bgroup\noexpand\next \let\next\f@@t
  \else\let\next\f@t\fi \next}
\def\f@@t{\bgroup\aftergroup\@foot\let\next}
\def\f@t#1{#1\@foot}
\def\@foot{\strut\egroup}
\def\footstrut{\vbox to\splittopskip{}}
\skip\footins=\bigskipamount 
\count\footins=1000 
\dimen\footins=8in 
   \def \bfootax  {\bgroup\tenrm
                  \baselineskip=12pt\lineskiplimit=-6pt
                  \hsize=19.5cc
                  \def\textindent##1{\hang\noindent\hbox
                  to\parindent{##1\hss}\ignorespaces}%
                  \footnotea{$^\star$}\bgroup}
   \def \efootax  {\egroup\egroup}
   \def \bfootay  {\bgroup\tenrm
                  \baselineskip=12pt\lineskiplimit=-6pt
                  \hsize=19.5cc
                  \def\textindent##1{\hang\noindent\hbox
                  to\parindent{##1\hss}\ignorespaces}%
                  \footnotea{$^{\star\star}$}\bgroup}
   \def \efootay  {\egroup\egroup }
   \def \bfootaz {\bgroup\tenrm
                  \baselineskip=12pt\lineskiplimit=-6pt
                  \hsize=19.5cc
                  \def\textindent##1{\hang\noindent\hbox
                  to\parindent{##1\hss}\ignorespaces}%
                 \footnotea{$^{\star\star\star}$}\bgroup}
   \def \efootaz {\egroup \egroup}
\def\fonote#1{\mehrsterne$^{\the\sterne}$\begingroup
       \def\textindent##1{\hang\noindent\hbox
       to\parindent{##1\hss}\ignorespaces}%
\vfootnote{$^{\the\sterne}$}{#1}\endgroup}
\catcode`\@=12
\everypar={\let\lasttitle=N\everypar={\parindent=1.5em}}%
%
%
\def \titlea#1{\stepc{Tl}
     \resetcount{Tm}
     \if A\lasttitle\else\vskip22pt\fi
     \ifdim\pagetotal>\pagegoal\else
     \setbox0=\vbox{
     \noindent\bf
     \rightskip 0pt plus4em
     \pretolerance=20000
     \arabic{Tl}.\
     \ignorespaces#1
     \vskip11pt}
     \dimen0=\ht0\advance\dimen0 by\dp0\advance\dimen0 by 4\baselineskip
     \advance\dimen0 by\pagetotal
     \ifdim\dimen0>\pagegoal\eject\fi\fi
     \bgroup
     \noindent
     \bf
     \rightskip 0pt plus4em
     \pretolerance=20000
     \arabic{Tl}.\
     \ignorespaces#1
     \vskip11pt
     \egroup
     \nobreak
     \parindent=0pt
     \everypar={\parindent=1.5em
     \let\lasttitle=N\everypar={\let\lasttitle=N}}%
     \let\lasttitle=A%
     \ignorespaces}
 \def\titleb#1{\stepc{Tm}
     \resetcount{Tn}
     \if N\lasttitle\else\vskip-11pt\vskip-\baselineskip \fi
     \vskip17pt
     \ifdim\pagetotal>\pagegoal\else
     \setbox0=\vbox{
     \raggedright
     \pretolerance=10000
     \noindent
     {\bf\arabic{Tl}}.\arabic{Tm}.\
     \ignorespaces#1
     \vskip8pt}
     \dimen0=\ht0\advance\dimen0 by\dp0\advance\dimen0 by 4\baselineskip
     \advance\dimen0 by\pagetotal
     \ifdim\dimen0>\pagegoal\eject\fi\fi
     \bgroup
     \raggedright
     \noindent \pretolerance=10000
     {\bf\arabic{Tl}}.\arabic{Tm}.\
     \ignorespaces#1
     \vskip8pt
     \egroup
     \nobreak
     \let\lasttitle=B%
     \parindent=0pt
     \everypar={\parindent=1.5em
     \let\lasttitle=N\everypar={\let\lasttitle=N}}%
     \ignorespaces}
 \def \titlec#1{\stepc{Tn}
     \resetcount{To}
     \if N\lasttitle\else\vskip-3pt\vskip-\baselineskip\fi
     \vskip11pt
     \ifdim\pagetotal>\pagegoal\else
     \setbox0=\vbox{
     \noindent
     \raggedright
     \pretolerance=10000
     {\bf\arabic{Tl}}.\arabic{Tm}.\arabic{Tn}.\
     {\it\ignorespaces#1}\vskip6pt}
     \dimen0=\ht0\advance\dimen0 by\dp0\advance\dimen0 by 4\baselineskip
     \advance\dimen0 by\pagetotal
     \ifdim\dimen0>\pagegoal\eject\fi\fi
     \bgroup
     \raggedright \noindent
     \pretolerance=10000
     {\bf\arabic{Tl}}.\arabic{Tm}.\arabic{Tn}.\
     {\it\ignorespaces#1}\vskip6pt
     \egroup
     \nobreak
     \let\lasttitle=C%
     \parindent=0pt
     \everypar={\parindent=1.5em
     \let\lasttitle=N\everypar={\let\lasttitle=N}}%
     \ignorespaces}
 \def\titled#1{\stepc{To}
     \resetcount{Tp}
     \if N\lasttitle\else\vskip-3pt\vskip-\baselineskip
     \fi
     \vskip 11pt
     \bgroup
     \it
     \noindent
     \ignorespaces#1\unskip. \egroup
     \let\lasttitle=N\ignorespaces}
\let\REFEREE=N
\newbox\refereebox
\setbox\refereebox=\vbox
to0pt{\vskip0.5cm\fullline{\hrulefill\tentt\lower0.5ex
\hbox{\kern5pt referee's copy\kern5pt}\hrulefill}\vss}%
\def\refereelayout{\let\REFEREE=M\footline={\copy\refereebox}%
\message{|A referee's copy will be produced}\par
\if N\lr\else
\if R\lr
\shipout\vbox{\makeheadline
\line{\box\leftcolumn}\makefootline}\advancepageno
\fi\let\lr=N
\topskip=10pt
\output={\plainoutput}%
\fi
}

\newcount\sterne \sterne=0
\newdimen\fullhead
\newtoks\RECDATE
\newtoks\ACCDATE
\newtoks\MAINTITLE
\newtoks\SUBTITLE
\newtoks\AUTHOR
\newtoks\INSTITUTE
\newtoks\SUMMARY
\newtoks\KEYWORDS
\newtoks\THESAURUS
\newtoks\SENDOFF
\newlinechar=`\|
\catcode`\@=\active
\let\INS=N%
\def@#1{\if N\INS $^{#1}$\else\if
E\INS\hangindent0.5\parindent\hangafter=1%
\noindent\hbox to0.5\parindent{$^{#1}$\hfil}\let\INS=Y\ignorespaces
\else\par\hangindent0.5\parindent\hangafter=1
\noindent\hbox to0.5\parindent{$^{#1}$\hfil}\ignorespaces\fi\fi}%
\def\mehrsterne{\advance\sterne by1\global\sterne=\sterne}%
\def\FOOTNOTE#1{\mehrsterne\ifcase\sterne
\or\bfootax \ignorespaces #1\efootax
\or\bfootay \ignorespaces #1\efootay
\or\bfootaz \ignorespaces #1\efootaz\else\fi}%
\def\PRESADD#1{\mehrsterne\ifcase\sterne
\or\bfootax Present address: #1\efootax
\or\bfootay Present address: #1\efootay
\or\bfootaz Present address: #1\efootaz\else\fi}%
\def\maketitle{\paglay%
\def\missing{ ????? }
%
\setbox0=\vbox{\parskip=0pt\hsize=\fullhsize\null\vskip2truecm
\edef\test{\the\MAINTITLE}%
\ifx\test\missing\MAINTITLE={MAINTITLE should be given}\fi
\aTa\ignorespaces\the\MAINTITLE\eTa
\edef\test{\the\SUBTITLE}%
\ifx\test\missing\else\aTb\ignorespaces\the\SUBTITLE\eTb\fi
\edef\test{\the\AUTHOR}%
\ifx\test\missing
\AUTHOR={Name(s) and initial(s) of author(s) should be given}\fi
{\tenpoint\leftskip=40pt\ignorespaces\noindent\the\AUTHOR\vskip5.7pt}
\let\INS=E%
\edef\test{\the\INSTITUTE}%
\ifx\test\missing
\INSTITUTE={Address(es) of author(s) should be given.}\fi
{\ninepoint\leftskip=40pt\ignorespaces\noindent\the\INSTITUTE\vskip11.4pt}%
\edef\test{\the\RECDATE}%
\ifx\test\missing
\RECDATE={$[$the date should be inserted later$]$}\fi
\edef\test{\the\ACCDATE}%
\ifx\test\missing
\ACCDATE={\begpet $[$the date should be inserted later$]$}\fi
{\leftskip=40pt\ninepoint\sl\noindent Received
\ignorespaces\the\RECDATE\unskip; accepted \ignorespaces
\the\ACCDATE\vskip20pt}%
\edef\test{\the\SUMMARY}%
\ifx\test\missing
\SUMMARY={Not yet given}\fi
{\ninepoint\leftskip=40pt\noindent{\bf Abstract{\kern.1em}.}\ \ ---\ \
\ignorespaces\the\SUMMARY\vskip11.4pt}
\edef\test{\the\KEYWORDS}%
\ifx\test\missing
\KEYWORDS={Key words should be inserted}\fi
{\ninepoint\leftskip=40pt\noindent{\bf Key words: }
\ignorespaces\the\KEYWORDS\vskip30pt \ \ }} 
\global\fullhead=\ht0\global\advance\fullhead by\dp0
\global\advance\fullhead by12pt\global\sterne=0
{\parskip=0pt\hsize=19.5cc\null\vskip2truecm
\edef\test{\the\SENDOFF}%
\ifx\test\missing\else\insert\footins{\smallskip\noindent\ninepoint
{\it Send offprint requests to\/}: \ignorespaces\the\SENDOFF}\fi
\hsize=\fullhsize
\edef\test{\the\MAINTITLE}%
\ifx\test\missing\message{|Your MAINTITLE is missing.}%
\MAINTITLE={MAINTITLE should be given}\fi
\aTa\ignorespaces\the\MAINTITLE\eTa
\edef\test{\the\SUBTITLE}%
\ifx\test\missing\message{|The SUBTITLE is optional.}%
\else\aTb\ignorespaces\the\SUBTITLE\eTb\fi
\edef\test{\the\AUTHOR}%
\ifx\test\missing\message{|Name(s) and initial(s) of author(s) missing.}%
\AUTHOR={Name(s) and initial(s) of author(s) should be given}\fi
{\tenpoint\leftskip=40pt\ignorespaces\noindent\the\AUTHOR\vskip5.7pt}
\let\INS=E%
\edef\test{\the\INSTITUTE}%
\ifx\test\missing\message{|Address(es) of author(s) missing.}%
\INSTITUTE={Address(es) of author(s) should be given.}\fi
{\ninepoint\leftskip=40pt\ignorespaces\noindent\the\INSTITUTE\vskip11.4pt}%
\edef\test{\the\RECDATE}%
\ifx\test\missing\message{|The date of receipt should be inserted
later.}%
\RECDATE={$[$the date should be inserted later$]$}\fi
\edef\test{\the\ACCDATE}%
\ifx\test\missing\message{|The date of acceptance should be inserted
later.}%
\ACCDATE={$[$the date should be inserted later$]$}\fi
{\leftskip=40pt\ninepoint\sl
\noindent Received \ignorespaces\the\RECDATE\unskip; accepted \ignorespaces
\the\ACCDATE\vskip20pt}%
\edef\test{\the\SUMMARY}%
\ifx\test\missing\message{|There is no Summary.}%
\SUMMARY={Not yet given.}\fi
{\ninepoint\leftskip=40pt\noindent{\bf Abstract{\kern.1em}.}\ \ ---\ \
\ignorespaces\the\SUMMARY\vskip11.4pt}
\edef\test{\the\KEYWORDS}%
\ifx\test\missing\message{|Missing keywords.}%
\KEYWORDS={Not yet given.}\fi
{\ninepoint\leftskip=40pt\noindent{\bf Key words: }
\ignorespaces\the\KEYWORDS\vskip30pt}} 

\edef\test{\the\THESAURUS}%
\ifx\test\missing\THESAURUS={missing; you have not inserted them}%
\message{|Thesaurus codes are not given.}\fi
\if M\REFEREE\let\REFEREE=Y
\normalbaselineskip=2\normalbaselineskip
\normallineskip=2\normallineskip\normalbaselines\fi
\global\sterne=0
\catcode`\@=12
\tenpoint
\let\lasttitle=A}

\MAINTITLE={A New Automatic Identification Technique for OB Associations
in Unresolved Galaxies}


 \AUTHOR={S. Adanti@1, P. Battinelli@2, R. Capuzzo--Dolcetta@1 and
P.W. Hodge@3 }

 \INSTITUTE={
 @1 Istituto Astronomico, Universita' di Roma La Sapienza, via G.M. Lancisi
 29, I-00161 Roma, Italy
 @2 Osservatorio Astronomico di Roma, viale del Parco Mellini 84, I-00136
 Roma, Italy
 @3 Astronomy Department, FM-20, University of Washington, Seattle,
 Washington 98195 }
\RECDATE={ XX-XX-19XX }
\ACCDATE={ XX-XX-19XX }
\SUMMARY={
	We present a new automatic technique based on Principal Component
Analysis and Cluster Analysis, with the aim of its application to the
identification
of those clumps in unresolved galaxies which likely represent regions
of star formation. We test the method by applying it to the galaxy M~31, for
which there are already several sets of identifications of OB associations
based on multi--colour images of resolved stars. We use small--scale digital
images of M~31 and compare the associations that we detect from these
unresolved data with previously--published large--scale data, finding a
rather good
agreement. We obtain a strict agreement of our identification
with the most compact associations of the
original van den Bergh (1964) identification.
We then apply the technique
to CCD images of the more distant spiral galaxy NGC~2903 and identify 68
OB association candidates.
}
\KEYWORDS={ Methods: data analysis -- galaxies: individual: M~31 -- galaxies:
 individual: NGC~2903 -- galaxies: star clusters }    
\SENDOFF={ R. Capuzzo--Dolcetta}          
\maketitle
\titlea {Introduction}

In recent years great effort has been expended to develop algorithms
suitable for an objective identification of structures in astronomical
images.
An objective method of identification of OB associations in galaxies is
important because the knowledge of the sites
and the sizes of star forming regions allows
inferences about the role of the environment in the mode of
star formation. Moreover, only an objective method allows
meaningful comparisons of OB associations in different parts of
a galaxy or in different galaxies.
It is well known that the identification by eye is
often unreliable (e.g. Hodge 1986), even in resolved fields, due to
biases introduced by
different observational characteristics and the subjective criteria used.
 For resolved galaxies objective methods were used successfully to identify
OB associations, for example, in the SMC (Battinelli 1991), in M~33
(Wilson 1991) and in M~31 (Magnier et al. 1993).
\par The aim of this paper is to present a  technique suitable
for objectively selecting OB associations in unresolved galaxies
(\ie\ galaxies not resolved into stars), using
photometric data only. We propose a method based on
a combined application of Principal Component Analysis (PCA) and
Cluster Analysis (CA) to fluxes and colours (Sect. 2).
To check our technique we decided to choose as a template the OB
associations of M~31, for which reliable identifications, based on resolved
images, are available. The comparison was thus made applying our technique
to images of M~31 whose resolution is low enough to simulate a distant
galaxy (Sect. 3).
A first application of the method is to NGC~2903 (Sect. 4).
\titlea {Methods and algorithms}

Let us suppose we have to analyse {\it n} objects
characterized by {\it s} variables (f$_1$, f$_2$,...,f$_s$).
 Geometrically, they can be seen as a cloud of points in a
real {\it s}-dimensional space {\bf $\Re^s$}.
The distance between any pair of items can be taken as
a criterion to see how much they look like each other, yielding
a subdivision into Clusters.
This clusterization requires the solution of
two problems:
 ({\it i}) distances are strongly biased by the different metrics
used and ({\it ii}) the high dimension $n\times s$ of the problem.
The usual way to overcome these problems is to define a new
space $\Re^p$ ($p<s$) in which ({\it i}) the distances are "homogeneous", and
({\it ii}) the $p$  ($p<s$) axes are comprehensive descriptors of variables.
Principal Component Analysis is the way to accomplish this task.

\titleb{P{\sc RINCIPAL COMPONENT ANALYSIS}}
{}From a mathematical point of view, the data
coming from {\it s} different CCD frames can be arranged into a
 bi-dimensional ${n\times s}$ array, \ie\ into a  matrix
A where {\it s} variables (columns) cross {\it n} objects (rows).
To make possible a meaningful comparison of the non--homogeneous
variables in A, it is usual to convert A into a matrix X where the
variables have zero mean, and variance is equal to one.
 Defining a new reference frame for the elements of the matrix X
as the one characterized by axes of
progressively decreasing variance (without changing the metric),
we can reduce the problem to
a space $\Re^p$ ($p<s$) defined by the first $p$ axes (factors)
which contain a prescribed amount of information (for example,
at least $90\%$ of the total variance). The space of the variables
with this reference frame is called Factorial Space.

\titleb{C{\sc LUSTER ANALYSIS}}
We perform the non-hierarchical
CA technique developed by Diday (1971).
The algorithm
resorts to the decomposition of the total inertia $I$ into
the sum of the  {\it "Intra-class Inertia"} and the {\it "Inter-class
Inertia"}.
Since $I$ is constant, minimizing Intra--class inertia
is equivalent to maximizing Inter--class inertia; in other words:
the more unlike the groups, the more
homogeneous each group is. The algorithm is iterative:
starting with an arbitrary choice of the initial partition
of the $n$ objects into a fixed number $m$ of classes, it should
be repeated  until the minimization of the Intra--class inertia is
achieved.\par
We checked the convergence of this method and  found  that after 50 iterations
the relative variation of the Intra--class inertia falls below 0.01.
In spite of this quick convergence, we noted that the final partition
into classes depends on the choice of the initial partition. This unpleasant
feature seems to be intrinsic in the method when $m \ll n$ and can be overcome
by choosing
an initial partition of $m' \gg m$ classes and then reducing the
number of classes in the final partition
from $m'$ to $m$, through a merging procedure.
The merging is an iterative procedure based on the distance
(in the space of the variables considered)
between the centroids of the classes: at each of the $m' - m$ steps needed,
 the two "nearest" classes are merged and the new centroid is computed.
 For all the CA applications
we performed,
no evident differences were found among final $m$--class partitions obtained
from different choices of
the initial $m'$--class partition. This makes us confident about the goodness
of the CA method when complemented with the merging procedure.
\par
Since $m$ is a free parameter of the CA, we found {\it a posteriori\/}
that $m=20$ can be considered a suitable choice for the kind of applications
proposed in this paper, \ie\  $m$ is large enough to account for all the
meaningful
structures in the frame but not too large to artificially split them.\par
The Fortran code to accomplish the PCA and CA steps of our technique has
been derived from Lebart \etal\ (1977).
\titlea {A check of the method with M 31}

We have checked our method using data for M~31 for which several
good independent
identifications of OB associations using single stars are available
(van den Bergh 1964, Efremov, Ivanov \& Nikolov 1987, Battinelli 1992,
Magnier \etal\ 1993).
\par We used photographic U,B,V,R plates of M~31 taken at the KPNO
with the Burrell--Schmidt 94-cm telescope, kindly provided by R.
Walterbos in digital form (see Walterbos \& Kennicutt 1987).
The whole M~31 field was scanned into 2 images (for each band)
composed by $2048\times 2048$
{\it pixels\/} ($1~pixel=~40 \mu m$); in order to have a resolution
comparable with realistic cases of unresolved galaxies, we preferred to
use images obtained by doubling the original pixel size, i.e. two
$1024\times 1024$ images. The corresponding scale is $7.73~arcsec/pixel$,
equivalent to $27~pc/pixel$ when a distance of $710~Kpc$, (Welch \etal\
1986) is assumed for M~31; note that $27~pc/pixel$ is almost the same scale
attainable if M~31 were $10~Mpc$ distant and observed with the KPNO 4-m
telescope. No removal of foreground stars has been done.
\par Due to the large computer storage required by the Cluster Analysis
code, we split each image into four $512\times 512$ fields.
The 4 outer fields
do not show any relevant structure
in the galaxy; this is likely due to the noise present in these
faint zones. Our attention is consequently devoted to just the 4
inner fields.
\par The application of PCA to M~31 flux and colour images shows that
it is sufficient to consider the first 4 factors
to account for $90\%$ of the total variance. Due to the high
correlation found among the U,B,V, and R fluxes, the choice of just one
flux and three colours causes a negligible loss of information.
Because of the dependence of CA on the metric of the reference frame only
(and not on the choice of the axes) for convenience we decided to resort
to the original standardized variables. Actually, this presents a
more straightforward interpretation of the CA results, and thus
we performed Cluster Analysis with 4 variables (3 colours:
U--B, B--V, V--R, and the flux U). However, we noted that the presence of the
flux leads to a classification where a large number of classes merely follows
the strong radial gradient of this variable present in the 4 inner fields.
This fact suggested to us to consider
only the 3 colour indexes, which actually lead to classifications
that more clearly delineate OB associations.\par
\begfig 11 cm
\figure{1} {Spatial distribution of our OB association candidates in Field 1
of M~31. Coordinates are in pixels. The contours overposed to the distribution
of the clumps are the outlines of the associations identified by van den
Bergh (1964).}
\endfig
Each application of our CA code leads to a partition
of the points in the space of the variables into 20 classes
corresponding to a partition on the galaxy frame. In this way, the whole result
of CA can be summarized into an "artificial" image whose pixels are
labelled by the class $(1-20)$ to which they belong.
OB association candidates were identified as
those classes whose pixels present a
knotty spatial distribution on the frame. Among
the "knotty" classes we then excluded those that clearly correspond
 to bright foreground stars leaving (generally)  one or two classes  as
association candidates.
It is worth noting that, as a direct effect of the noise in the data,
the classes we get often contain isolated pixels which are not related
to any meaningful structure. In order to represent better the real features
in the  class of the knots, we  removed "isolated" pixels using the PLC
procedure (Battinelli 1991, Battinelli \& Demers 1992) with a threshold
of $90\%$ of statistical significance.\par
     Even though more recent identifications of associations in M~31 based
on high resolution observations have
been published,
	we preferred to compare our results with the van den Bergh (1964)
(hereafter vdB) identification whose image resolution is more similar
to ours. In Figures 1--4 our OB association candidates in the inner
fields of M~31 are displayed together with the contour--lines of vdB's
associations.\par
\titleb{F{\sc IELD 1}}
This field (East position) is characterized by a large
number of dense and outstanding associations (see Figure 2 in vdB).
Most of them,
as described by van den Bergh, show dense sub--structures which, more recently,
have been recognized as "aggregates" and "associations"  by Efremov \etal\
(1987).\par
 Using our procedure we were able to select association candidates
that are in fairly good agreement with vdB's (see Figure 1).
Indeed, we found a
total of 39 clumps; 32 of them are spatially related to vdB
associations; 4 out of the 7 clumps without vdB counterpart correspond
to 4 "star complexes" identified by Efremov \etal\ (1987).
Of a total of 33 vdB objects of this field, we missed 13;
only two or three of these associations show dense sub--structures.
The clump related to the largest association of the field, vdB
OB 48, shows the presence of sub--clumping, in agreement with vdB.\par
	This is the field  that yields the best identification of our clumps
with van den Bergh's associations: this is due to the fact that the low
resolution of our images allows our method to identify mainly dense and well
defined (from the surrounding regions) clumps, and this field is particularly
rich in compact groupings of bright stars.\par
\begfig 11 cm
\figure{2} {Field 2 of M~31. For symbols see Figure 1.}
\endfig
\begfig 11 cm
\figure{3} {Field 3 of M~31. For symbols see Figure 1.}
\endfig
\titleb{F{\sc IELD 2}}
Even if the number of associations (32) found by {\nobreak vdB} in this Field
is
about the same as in Field 1, these associations appear intrinsically
different.
Indeed, vdB's associations in Field 2 (South position) mainly consist of
large regions which (with the exception of a few cases) are not well separated
from the background, and, moreover, intrinsically lack a real clumpy
structure, as confirmed by recent high resolution data (see for instance
Fig. 3 of Magnier \etal\ 1993).\par
When applied to this field, our procedure yields classes that describe
sufficiently well the spiral arm pattern, but fails to make out an evident
knotty structure along the arm, in agreement with the characteristics of
the field; only a few of our clumps are related to vdB associations (see
Figure 2). More precisely, we find 21 association candidates: 12 of them
have a vdB counterpart; 3 of the remaining 9 have been identified
as "complexes" and 2 as "regions with enhanced density of B--stars" by
Efremov \etal\ (1987).
\titleb{F{\sc IELD 3}}
This field (West position) has similar characteristics to those
of Field 2; that is, the
vdB associations have in only very few cases a clumpy structure
emerging well above the background.
Thus it is not surprising that in this field all
the classes (with the exception of foreground stars) found by our procedure
have a diffuse aspect, so it is difficult to select a really "knotty" class.
However, the less diffuse class traces well the spiral structure. In
Figure 3 we show the distribution of the statistically significant
($90\%$) clumps inside the class. A part from the well recognizable NGC 206,
their positions do not agree with the positions of vdB's associations.
\titleb{F{\sc IELD 4}}
The associations of this field (North position) are well separated from the
surrounding
environment, but (with the exception of vdB OB 54) they are not as dense
as those of Field 1.
	Using our procedure, we found 27 clumps, 21 of which are related to
vdB objects; only two of the remaining 6 clumps are classified by
Efremov \etal\ (1987) as a "star complex" and a "region with enhanced
density of B--stars".
\begfig 11 cm
\figure{4} {Field 4 of M~31. For symbols see Figure 1.}
\endfig
{\noindent We were able to identify 14 out of the 38 van den Bergh's
associations (see Figure 4).
It is worth noting that most of these associations have a dense
core, while almost all of the vdB objects we missed are low--density,
scattered
associations of stars.}\par
\vskip 12pt
	Thus, the application of our methodology to low resolution
images
of M~31 allows a fairly good identification of OB associations. The good
agreement for Fields 1 and 4, together with the poor agreement for Fields
2 and 3, suggests
that this technique
is able to identify mainly the dense sub--structures of van den Bergh's
objects. Many of the clumps that we identify and do not correspond to a
vdB association have a counterpart in the Efremov \etal\ (1987) classification.
It is difficult (it requires a specific analysis) to state firmly about the
remaining (not previously identified as associations) few clumps. Indeed, they
might be real OB associations, or regions of low absorption. Taking into
account that our aim to apply the methodology to unresolved fields
in distant galaxies suggested us to choose data of poor resolution for
M~31, we retain the results discussed here as strongly encouraging to
an application to a more distant galaxy.
\titlea {An application to NGC 2903}

Data used for our first application of this identification technique are
CCD frames with $512\times 320$ pixels centered on 3 different positions
in the galaxy NGC~2903. They cover the whole galaxy with some overlap.
The frames were taken with the KPNO 4-m telescope, using UBVR filters
(Hodge, Jaderlund \& Meakes 1990).
Each image covers a $5' \times 3'$ area; the scale is $0.59~arcsec/pixel$,
corresponding -- at the distance of NGC~2903 (about 10 Mpc, Sandage
and Tammann 1981) -- to $\sim 29~pc/pixel$.\par
NGC~2903 is an Sc spiral of luminosity class I-II. Its central region,
which has been extensively studied (Sersic \& Pastoriza 1967; Oka
\etal\ 1974; and others), shows bright
\lq hot spots \rq\ on a scale of the order of $20~arcsec$ in an area of
active star formation. As indicated by Wevers (1984) and Hodge \etal\
(1990), NGC~2903 shows luminosity and colour profiles
suggesting active star formation throughout its spiral
structure. The outer north--east (NE) arm shows a particularly
well defined distinction between a smooth and diffuse component and
one showing bright knots, interpreted as sites of
recent star formation.
\titleb{N{\sc ORTH--EAST FIELD}}
\par The first application of our CA
methodology was to the outer NE UBVR frames (Field 4 in Hodge \etal\ 1990).
They were used to compute U--B, B--V and V--R images, after sky background
subtraction; consequently the matrix to be handled consists of
$(512\times 320)\times 7$ entries.
PCA analysis allowed us to restrict ourselves to one flux
and three colour frames, leading to a $40\%$ reduction in
the data. The partition in classes leads to a clear distinction
between a diffuse and a knotty arm. The knotty arm
(which was already identified and discussed by Hodge \etal\ 1990)
is defined by 4 classes. Anyway, it is clear that one class is sufficient
to identify the knots well (see Figure 5).
The remaining 3 classes correspond to the innermost high luminosity regions.
Note that Figure 5,
as well as Figures sketching the other NGC~2903 fields, were obtained
through the PLC procedure mentioned in the previous Section.
This filtering had a small effect on the clumps, which were already
well defined.
In Table 1 we give values of the clump
sizes, calculated as the average between their x and y extents. They seem
to be fairly large compared to typical association dimensions.
The clump size values can be considered as overestimates,
due to the seeing ($1.8~arcsec$, corresponding to $\sim 3$ pixels), and
because many of them may consist of unresolved aggregates
of more than one association.
In the same Table we also present the surface brightness in the B band, the
absolute B magnitudes and the (U--B)$_0$, (B--V)$_0$ and (V--R)$_0$ colours
of the selected clumps.
\begfig 11 cm
\figure{5} {The North--East Field of NGC~2903. Each clump is one of our
OB association candidates, and is labelled by its identification number in
this field (see also Tables 1--3). Coordinates are in pixels.}
\endfig
\input table1.tex
\input table2.tex
{\noindent The reddening correction is taken from
de Vaucouleurs \etal\ (1991) and accounts only for Galactic
extinction, $A_{g}=0.07~mag$, and the average component of the internal
NGC~2903 extinction $A_{i}=0.47~mag$ in the B-band. Of course, this
probably underestimates the reddening in the listed areas, which likely
have more dust than the average.}
\begfig 11 cm
\figure{6} {The South--West Field of NGC~2903. Symbols and numbers as in
Figure 5.}
\endfig
\titleb{S{\sc OUTH--WEST FIELD}}
One of the classes of the partition shows some
knots embedded in a diffuse arm in contrast to the North--East Field, where
the diffuse and knotty arms are well separated. The clumps of this
field (see Figure 6) are smaller than in the other fields, as
shown in Table 2.
\titleb{I{\sc NNER FIELD}}
This field corresponds to the central region of the galaxy, which is,
of course,
a high luminosity region. The class partition is able to outline
the inner spiral structure.
As in the North--East Field,
one class suffices to outline some blue (high U/B flux ratio) and bright
clumps (see Figure~7), which we identify as regions of
star formation. Their characteristics are given in Table 3. The clump
colours are significantly redder than those of the other fields; this
seems to indicate a high local absorption, as qualitatively suggested
by Figure 8.
\input table3.tex
{\noindent Note that while the centroids of the South--West and Inner
Fields lie on the same reddening
line, the North--East Field is
intrinsically bluer, thus indicating earlier spectral types.
The mean $A_{\rm B}$ absorptions obtained by the reddening lines
are $\sim$ 2.4, 2.3 and 3.8 $mag$ for the North--East, South--West
and Inner Fields, respectively.}\par
The low  resolution of this image together with saturation of the brightest
areas in some frames do not allow us to identify the 'hot spots' previously
identified in the galaxy centre by other authors.
\vskip 12 pt
\begfig 11 cm
\figure{7} {The Inner Field of NGC~2903. Symbols and numbers as in
Figure 5.}
\endfig
	It is worth noting that our determined clump distributions show
good correspondence to the peaks in the (U--B) colour maps (see
Figures 9, 10 and 11).
\begfig 8. cm
\figure{8} {Colour--colour diagram for the OB association candidates found
in the North--East Field (triangles), South--West Field (squares) and
Inner Field (circles) of NGC~2903. The standard Main Sequence (solid line)
is also plotted (Johnson 1966). The straight line is the reddening line for
the Inner and South--West clumps.}
\endfig
\begfig 9.5 cm
\figure{9} {Colour map of the North--East Field in (U--B) overposed to the
distribution of our clumps. Contour levels are --0.4 and --0.5.}
\endfig
\begfig 9. cm
\figure{10} {Colour map of the South--West Field in (U--B) overposed to the
distribution of our clumps. Contour levels are 0.0 and --0.1.}
\endfig
\begfig 9.5 cm
\figure{11} {Colour map of the Inner Field in (U--B) overposed to the
distribution of our clumps. Contour levels are 0.1 and 0.0.}
\endfig
\titlea {Conclusions}

In this paper we have presented and discussed a new objective
technique, based on Principal Component Analysis and Cluster Analysis, aiming
at identifying OB associations and other structures in unresolved
galaxies.
\par We have tested our method with M~31 data, comparing the OB association
identification obtained with our technique to previous identifications, finding
a sufficiently good agreement for bright, compact associations.
The first application of our method is to NGC~2903 CCD data.
Our classification is able to identify a total
of $68$ OB association candidates in the areas studied. A strict
overlap of the distribution of OB association candidates
with the bluest zones in the galaxy's colour map is found.
%
\ack{We would like to thank Dr. R. Walterbos for providing us with
his frames of M~31.}
%
\begref

\ref Battinelli P., 1991, \aa 244 69
\ref Battinelli P., 1992, \aa 258 269
\ref Battinelli P., Demers S., 1992, \aj 104 1458
\ref de Vaucouleurs G., de Vaucouleurs A., Corwin H.G., Buta R.J.,
Paturel G., Fouqu\'e P.,
 1991, {\it Third Reference Catalogue of
Bright Galaxies} (Springer-Verlag, New York)
\ref Diday E., 1971, {\it La M\'{e}thode des Nu\'{e}es Dynamiques}, Rev.
	Stat. App., vol. 19, n. 2, p. 19
\ref Efremov Yu.N., Ivanov G.R., Nikolov N.S., 1987, \apss 135 119
\ref Hodge P.W., 1986, in {\it Luminous Stars and Associations in Galaxies},
	IAU Symp. 116, eds. C.W.H. De Loore, A.J. Willis, P.G. Laskarides,
	Reidel, Dordrecht, p. 369
\ref Hodge P.W., Jaederlund E., Meakes M., 1990, \pasp 102 1263
\ref Johnson H. L., 1966, \arev 4 193
\ref Lebart L., Morineau A., Tabard N., 1977, {\it Techniques de la description
	statistique} (Dunod, Paris)
\ref Magnier E.A., Battinelli P., Lewin, W.H.G., Haiman Z.,
	van Paradijs J., Hasinger G., Pietsch W., Supper R.,
	Tr$\ddot{\rm u}$mper J., 1993, \aa 278 36
\ref Oka S., Wakamatsu K., Sakka K., Nishida M., Jugaku J.,
1974, \pasp 26 289
\ref Sandage A.R., Tammann G.A., 1981, {\it A Revised Shapley--Ames Catalog
of Bright Galaxies} (Carnegie Inst., Washington)
\ref Sersic J.L., Pastoriza M., 1967, \pasp 79 152
\ref van den Bergh S., 1964, \apjs 86 65
\ref Walterbos R.A.M., Kennicutt R.C., 1987, \aas 69 311
\ref Welch D.L.,  McAlary C.W., McLaren R.A., Madore B.F., 1986,
	\apj 305 583
\ref Wevers B.M.H.R., 1984,  Ph.D. Thesis, Gr\" oningen
\ref Wilson C.D., 1991, \aj 101 1663
\endref
\vfill\eject
\bye